\documentclass[aps,pre,twocolumn,showpacs,preprintnumbers,amsmath,amssymb,groupedaddress]{revtex4}
\usepackage{graphicx}
\usepackage{txfonts}

\begin{document}


\title{The Deviation of the Vacuum Refractive Index Induced by a Static Gravitational Field}


\author{Xing-Hao Ye, Qiang Lin}
\email[Electronic address:]{qlin@zju.edu.cn}

\affiliation{Department of Physics, Zhejiang University, Hangzhou
310027, China }


\date{\today}

\begin{abstract}
We analyzed the influence of static gravitational field on the
vacuum and proposed the concept of inhomogeneous vacuum. According
to the observational result of the light deflection in solar
gravitational field as well as the corresponding Fermat's principle
in the general relativity, we derived an analytical expression of
the refractive index of vacuum in a static gravitational field. We
found that the deviation of the vacuum refractive index is composed
of two parts: one is caused by the time dilation effect, the other
is caused by the length contraction effect. As an application, we
simulated the effect of the gravitational lensing through computer
programming and found that the missing central imaging could be
interpreted in a reasonable way.
\end{abstract}

\pacs{42.25.Bs, 42.50.Lc, 04.}

\maketitle


\section{Introduction}
\label{} Vacuum is usually considered as ``homogeneous'' and
``isotropic'', i.e., vacuum does not differ from place to place, and
the refractive index of vacuum is always equal to 1. However, the
recent theoretical and experimental progresses demonstrate that such
concept of vacuum turns out to be inappropriate when there are
matters or fields within finite distance. For example, the vacuum
inside a microcavity is modified due to the existence of the cavity
mirrors, which will alter the zero-point energy inside the cavity
and cause an attractive force between the two mirrors known as
Casimir effect \cite{rf-Gies2006,rf-Emig2006}, which has been
verified experimentally \cite{rf-Lamoreaux1997,rf-Chan2001}. A
second example is that, under the influence of electromagnetic
field, vacuum can be polarized, which has led to astonishingly
precise agreement between predicted and observed values of the
electron magnetic moment and Lamb shift, and may influence the
motion of photons \cite{rf-Ahmadi2006}. Dupays \emph{et al}.
\cite{rf-Dupays2005} studied the propagation of light in the
neighborhood of magnetized neutron stars. They pointed out that the
light emitted by background astronomical objects will be deviated
due to the optical properties of quantum vacuum in the presence of a
magnetic field. Also in \cite{rf-Rikken2003}, Rikken and Rizzo
considered the anisotropy of the optical properties of the vacuum
when a static magnetic field $\textbf{B}_0$ and a static electric
field $\textbf{E}_0$ are simultaneously applied perpendicular to the
direction of light propagation. They predicted that magnetoelectric
birefringence will occur in vacuum under such conditions. They also
demonstrated that the propagation of light in vacuum becomes
anisotropic with the anisotropy in the refractive index being
proportional to ${\textbf{B}_0}\times{\textbf{E}_0}$.

The facts that the propagation of light in vacuum can be modified by
applying electromagnetic fields to the vacuum implies that the
vacuum is actually a special kind of optical medium
\cite{rf-Ahmadi2006,rf-Dupays2005}. This is similar to the Kerr
electro-optic effect and the Faraday magneto-optic effect in
nonlinear dielectric medium. This similarity between the vacuum and
the dielectric medium implies that vacuum must also have its inner
structure, which could be influenced by matter or fields as well.
Actually, the structure of quantum vacuum has already been
investigated in quite a number of papers
\cite{rf-Armoni2005,rf-Barroso2006,rf-Dienes2005}.

In this paper, with the analysis of the influence of static
gravitational field on the vacuum, we put forward a new concept that
the curved spacetime around a certain matter can be treated as an
optical medium with a graded refractive index. We suggest that the
so-called curved spacetime is a reflection of the vacuum
inhomogeneity caused by the influence of gravitational matter. Based
on this idea, the refractive index of vacuum is derived. We will
also apply this concept to unpuzzle the problem of the central image
missing in almost all the observed cases of gravitational lensing
\cite{rf-Winn2004}.

\section{The deviation of the vacuum refractive index}
\label{} According to the astronomical observation, the light
propagating through a space with a celestial body nearby will be
deflected. It can be interpreted with the curved spacetime in
general relativity. As a matter of fact, it can also be interpreted
with the assumption that the vacuum around matter is inhomogeneous
with refractive index deviated from 1. Here we put forward a
theoretical model to describe the refractive index profile based on
the Fermat's principle for the propagation of light in a static
gravitational field, which was given by Landau and Lifshitz
\cite{rf-Landau1975}:
\begin{eqnarray}
\delta \int {g_{00}}^{-1/2}dl=0,
\end{eqnarray}
where $dl$  is the local length element passed by light and measured
by the observer at position $r$ in the gravitational field, $r$ is
the distance from this element of light to the center of
gravitational matter $M$, $g_{00}$ is a component of the metric
tensor $g_{\mu\nu}$, $g_{00}^{-1/2}dl$ corresponds to an element of
optical path length. ${g_{00}}^{-1/2}=dt/d\tau$, where $d\tau$
represents the time interval measured by the local observer for a
light ray passing through the length $dl$, while $dt$ is the
corresponding time measured by the observer at infinity. Eq.(1)
could then be rewritten as
\begin{eqnarray}
\delta \int {g_{00}}^{-1/2}dl&=&\delta \int \frac{dt}{d\tau} dl\nonumber\\
&=&\delta \int \frac{dt}{d\tau}\frac{dl}{ds} ds\nonumber\\
&=&\delta \int n ds=0,
\end{eqnarray}
where $ds$ is the length element measured by the observer at
infinity, corresponding to the local length $dl$.

Eq.(2) shows that if we set the scale of length and time at infinity
as a standard scale for the whole gravitational space and time, the
propagation of light then satisfies the standard representation of
Fermat's principle, with the space --- actually the vacuum ---
possessing a refractive index given by
\begin{eqnarray}
n=\frac{dt}{d\tau} \frac{dl}{ds} = n_1  n_2.
\end{eqnarray}

The factor $n_1$ of the refractive index relating to the time
transformation effect $dt/d\tau$ can be derived from the Newtonian
attraction, which contributes partially to the deflection of light.
Considering a photon of relativistic mass $m_\infty$ at the infinity
moving down to position $r$, the work done on to the photon by the
Newtonian gravity is
\begin{eqnarray}
-\frac{GMm}{r^2} dr=d(mc^2),
\end{eqnarray}
where $G$ is the gravitational constant, $c$ is the velocity of
light, $M$ is the mass of a star (say the Sun), $r$ is the distance
to the center of the star. Integrating Eq.(4) gives
\begin{eqnarray}
m_r=m_\infty e^\frac{GM}{rc^2},
\end{eqnarray}
where $m_r$ is the relativistic mass of the photon at position $r$.

Since the photon energy is $E=h \nu=mc^2$, where $h$ is the Planck
constant, $\nu$ is the photon frequency, then we have $m=h \nu
/c^2$. Substituting it into Eq.(5) gives
\begin{eqnarray}
\nu_r=\nu_\infty e^\frac{GM}{rc^2}.
\end{eqnarray}
It is just the frequency shift caused by the gravitational force,
which reflects that a clock in a gravitational field runs slower
than that far away from the gravitational center. That is
\begin{eqnarray}
d\tau=e^{-\frac{GM}{rc^2}}dt,
\end{eqnarray}
where $d\tau$ denotes the time measured by a clock at position $r$,
$dt$ is the converted time of $d\tau$, i.e., the time measured by
the clock at infinity. This relation indicates that, if the length
scale is the same, i.e., $dl=ds$, when an observer at position $r$
reports a light velocity $c_1=dl/d\tau$, it should be converted by
the observer at infinity into
\begin{eqnarray}
c_1'=\frac{ds}{dt}=\frac{dl}{e^\frac{GM}{rc^2}d\tau }=c_1
e^{-\frac{GM}{rc^2}}.
\end{eqnarray}
This change of light velocity will certainly bring a deflection to
the light propagation. The corresponding refractive index is
\begin{eqnarray}
n_1=\frac{c_1}{c_1'}=\frac{dt}{d\tau}=e^\frac{GM}{rc^2}.
\end{eqnarray}

Let us now consider the deflection angle caused by this graded
refractive index. In Fig.1, the curve AP represents the light ray,
$\beta$ is the angle between the position vector $\textbf{r}$ and
the tangent at the point P on the ray, $\varphi$ is the deflection
angle of light. Since the refractive index shown in Eq.(9) has
spherical symmetry, i.e., depends only on the distance $r$ for a
given mass $M$, according to the Fermat's principle
\begin{figure}
\centering
\includegraphics{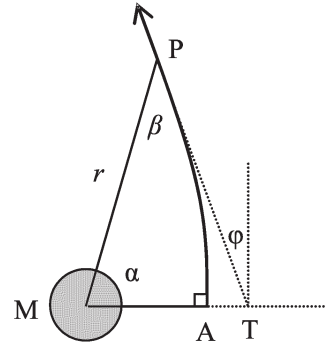}
\caption{\label{fig01} Light deflection caused by a graded
refractive index.}
\end{figure}
\begin{eqnarray}
\delta \int n ds=0,
\end{eqnarray}
where $ds=dr \sqrt{1+(r\dot{\alpha})^2}$, $\dot{\alpha}=d\alpha/dr$,
$n=n(r)$, we have the corresponding Lagrangian function
\begin{eqnarray}
L(\alpha,\dot{\alpha};r)=n(r) \sqrt{1+(r\dot{\alpha})^2}.
\end{eqnarray}
Using the Lagrangian equation
\begin{eqnarray}
\frac{d}{dr}(\frac{\partial L}{\partial
\dot{\alpha}})-\frac{\partial L}{\partial \alpha}=0,
\end{eqnarray}
we get \cite{rf-Wolf1999}
\begin{eqnarray}
n r \sin{\beta}=\textnormal{constant},
\end{eqnarray}
or
\begin{eqnarray}
n r \sin{\beta}=n_0 r_0,
\end{eqnarray}
where $r_0$ and $n_0$ represent the radius and refractive index at
the nearest point A respectively.

Since
\begin{eqnarray}
\tan{\beta}=\frac{rd\alpha}{dr},
\end{eqnarray}
associating with Eq.(14) reaches
\begin{eqnarray}
d\alpha=\frac{dr}{r \sqrt{(\frac{nr}{n_0 r_0})^2-1}}.
\end{eqnarray}
\begin{figure}
\centering
\includegraphics{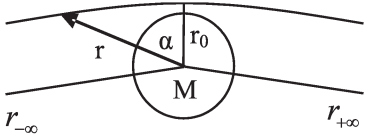}
\caption{\label{fig02} Light deflection in solar gravitational
field.}
\end{figure}

For a light ray passing by the Sun as shown in Fig.2, the total
angular displacement of the radius vector $\textbf{r}$ reads
\begin{eqnarray}
\Delta \alpha=2 \int_{r_0}^{\infty} \frac{dr}{r \sqrt{(\frac{nr}{n_0
r_0})^2-1}},
\end{eqnarray}
where $r_0$ represents the nearest distance to the center of the
Sun. Because the gravitational field of the Sun is a weak field, the
value of $GM/rc^2$ is quite small, so substituting Eq.(9) into
Eq.(17) gives a solution of first order approximation
\begin{eqnarray}
\Delta \alpha=\pi + \frac{2GM}{r_0 c^2}.
\end{eqnarray}

Then the total deflection angle of light caused by the refractive
index $n_1$ in solar gravitational field is
\begin{eqnarray}
\Delta \varphi_1 = \Delta \alpha - \pi = \frac{2GM}{r_0 c^2}.
\end{eqnarray}

In fact, this result was obtained early in 1911 by Einstein
\cite{rf-Einstein1923}, who also investigated the effect of red
shift and the corresponding slowing down of the light velocity in
gravitational field and then figured out the light deflection as
shown in Eq.(19) with the use of Huygens' principle. Since the
actual total deflection angle of light propagation calculated by the
general relativity \cite{rf-Ohanian1976,rf-Weinberg1972} and
measured by the astronomical observation \cite{rf-Fomaleont1976} is
twice that value, we then come to know that the length
transformation effect $dl/ds$ in Eq.(2) must have the same relation
as that of the time transformation effect $dt/d \tau$ expressed in
Eq.(9), namely
\begin{eqnarray}
\frac{dl}{ds}=e^\frac{GM}{rc^2}.
\end{eqnarray}

This relation indicates that a ruler in a gravitational field is
shorter than that far away from the gravitational center. So when an
observer at position $r$ reports a length $dl$, it should be
converted by the observer at infinity into
\begin{eqnarray}
ds=e^{-\frac{GM}{rc^2}}dl.
\end{eqnarray}

For a light passing through a length $dl$, if the time scale is the
same, i.e., $d \tau =dt$ , the light velocity $c_2=dl/d \tau$
reported by the observer at position $r$ should then be converted by
the observer at infinity into
\begin{eqnarray}
c_2'=\frac{ds}{dt}=\frac{e^{-\frac{GM}{rc^2}}dl}{d \tau}=c_2
e^{-\frac{GM}{rc^2}}.
\end{eqnarray}

This change of light velocity will also bring a deflection to the
light propagation. The refractive index corresponding to this kind
of deflection is
\begin{eqnarray}
n_2=\frac{c_2}{c_2'}=\frac{dl}{ds}=e^\frac{GM}{rc^2},
\end{eqnarray}
which also causes a deflection angle of light
\begin{eqnarray}
\Delta \varphi_2 = \frac{2GM}{r_0 c^2}.
\end{eqnarray}
Therefore, the total deflection angle of light in solar
gravitational field is
\begin{eqnarray}
\Delta \varphi =\Delta \varphi_1 + \Delta \varphi_2 = \frac{4GM}{r_0
c^2}.
\end{eqnarray}

The above result shows that, if the two refraction effects are
considered simultaneously, then the gravitational space --- actually
the vacuum in the gravitational field --- can be regarded as an
optical medium with a total refractive index given by
\begin{eqnarray}
n=n_1 n_2 = e^\frac{2GM}{rc^2}.
\end{eqnarray}
$n$ is composed of two factors: $n_1$ --- related with the time
transformation or ``curved time''; $n_2$ --- related with the space
transformation or ``curved space''. So the curved spacetime of
general relativity is reflected in the synthesized refractive index
$n$, which is also a reflection of the inhomogeneity of the vacuum,
showing that the vacuum near the matter is influenced more than that
far away from the matter.

\begin{figure}
\centering
\includegraphics[totalheight=1.7in]{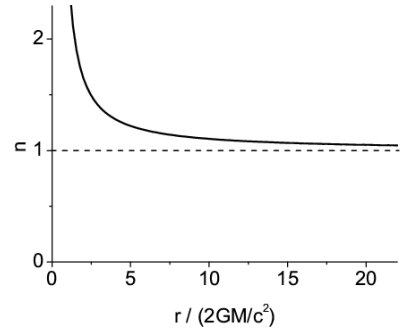}%
\caption{\label{fig03}The dependence of the vacuum refractive index
$n$ on the distance $r$.}
\end{figure}

The above expression of $n$ shows that the refractive index of the
vacuum at the infinity from the gravitational matter is 1, i.e., the
usual refractive index of vacuum. The closer of the position to the
center of matter $M$, the higher the refractive index of the vacuum.
The relation between $n$ and $r$ is depicted in Fig.3, where
$2GM/c^2$ is taken as the unit of $r$. For examples, the
corresponding radii for the surface of the Sun in solar
gravitational field and the surface of the Earth in earth
gravitational field are $2.36 \times 10^5$ and $7.20 \times 10^8$
respectively --- both are far beyond the $r$-axis illustrated in
Fig.3.

The deviation of the vacuum refractive index from the usual value 1
is given by
\begin{eqnarray}
\Delta n=n-1= e^\frac{2GM}{rc^2}-1.
\end{eqnarray}

In weak field it becomes
\begin{eqnarray}
\Delta n=\frac{2GM}{rc^2}.
\end{eqnarray}

In order to provide the readers with a quantity impression, let us
give two examples. For the solar gravitational field ($M=1.99 \times
10^{30} kg$), the deviation of $n$ on the surface of the Sun
($r=6.96 \times 10^8 m$) is $4.24 \times 10^{-6}$. For the earth
gravitational field ($M=5.98 \times 10^{24} kg$), the deviation of
$n$ on the surface of the Earth ($r=6.38 \times 10^6 m$) is only
$1.39 \times 10^{-9}$, which is so small that it can hardly be
observed in usual experiments. Nevertheless, for a massive celestial
body such as a heavy star, a galaxy or a cluster of galaxies, the
deviation is not only observable, but also important and useful in
gravitational astronomy.

\section{Applications}
\label{} The deflection of light by massive bodies leads to the
effect of gravitational lensing. Formerly, this effect should be
calculated complicatedly with the general relativity
\cite{rf-Mollerach2002}. Once we have introduced the concept of
graded vacuum refractive index and obtained its relation with mass
$M$ and position $r$, the problem of gravitational lensing could
then be treated easily with the conventional optical method.

Considering a source $S$ and a lens $L$ of mass $M$, the light
emitted from $S$ is bent due to the gravitational field of the lens.
The bent light could be figured out through Eq.(13) and Eq.(26).
Drawing the extension line of the light from the observer $O$, the
apparent (observed) position of the source image $I$ could then be
found out. The result is shown in Fig.4.

\begin{figure}
\centering
\includegraphics[totalheight=1.7in]{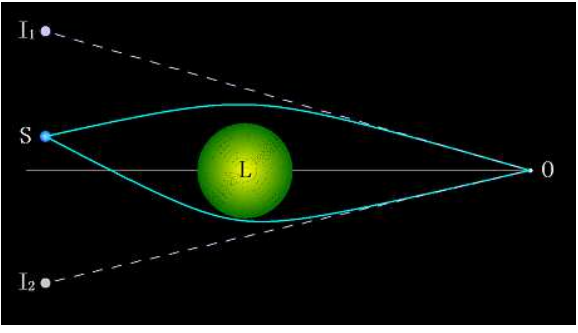}
\caption{\label{fig04}  The two images $I_1$, $I_2$ of a
gravitational lens.}
\end{figure}

This method could also be applied in studying the central imaging.
In doing this, the vacuum refractive index profile inside the
lensing body should be considered as well.

Noticing that Eq.(26) could be virtually rewritten as
\begin{eqnarray}
n= e^{-\frac{2P_r}{c^2}},
\end{eqnarray}

where $P_r$ represents the gravitational potential at position $r$
from the center of the lens.

As a model for discussion, we suppose a lens (for example, a galaxy
or a cluster of galaxies) of radius $R$ with a density distribution
\begin{eqnarray}
\rho=\rho_0[1-(\frac{r}{R})^k],
\end{eqnarray}
where $\rho_0$ is the central density of the lens, $0\leqslant
r\leqslant R$, $k>0$. The density $\rho$ decreases with the distance
$r$ from the center of mass; the decreasing varies with the
parameter $k$. This model gives the distribution of gravitational
potential as

\begin{eqnarray}
P_o=-4\pi \rho_0
G\frac{k}{3(3+k)}\frac{R^3}{r};\nonumber \\
P_i=-4\pi \rho_0 G
\{\frac{k}{2(2+k)}R^2-[\frac{1}{6}-\frac{1}{(2+k)(3+k)}(\frac{r}{R})^k]r^2\}
,
\end{eqnarray}
for outside ($r\geqslant R_0$) and inside ($r\leqslant R_0$) the
gravitational lens respectively.

The vacuum refractive index profile outside and inside the
gravitational lens then reads

\begin{eqnarray}
n_o=\textnormal{exp}\left[\frac{8 \pi \rho_0
G}{c^2}\frac{k}{3(3+k)}\frac{R^3}{r} \right];\nonumber \\
n_i=\textnormal{exp}\left\{\frac{8\pi \rho_0
G}{c^2}\{\frac{k}{2(2+k)}R^2-[\frac{1}{6}-\frac{1}{(2+k)(3+k)}(\frac{r}{R})^k]r^2\}\right\}.
\end{eqnarray}

\begin{figure}
\centering
\includegraphics[totalheight=1.7in]{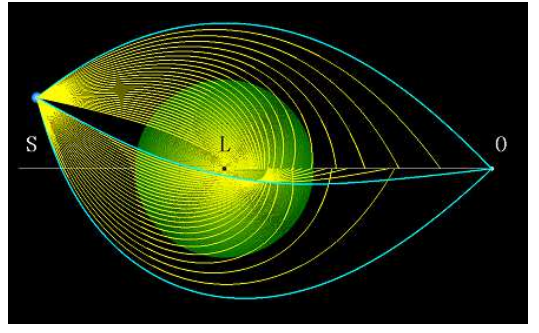}
\caption{\label{fig05}  A ray tracing result for the central
imaging.}
\end{figure}

Fig.5 shows a ray tracing result for the imaging of a gravitational
lens with the above described vacuum refractive index profile. In
the figure, only three paths of ray (the three thick lines) could
pass through the observer $O$, forming the upper, lower and central
images respectively. From the figure, we find that, under the same
conditions, the larger the distance $OL$ from the observer to the
lens, the closer the central imaging light to the center of the
lens. If the source $S$ and the observer $O$ are counterchanged, it
could also be known from the figure that, the larger the distance
$SL$ from the source to the lens, the closer the central imaging
light to the center of the lens. In addition, through the change of
the lens mass $M=\frac{4}{3}\pi R^3 \rho_0 k/(3+k)$, we also find
that, when the mass $M$ increases, the distance from the central
imaging light to the center of the lens decreases (Fig.6, where the
mass ratio of the lenses corresponding to the four central imaging
rays from bottom to top is $2:3:4:5$ ).

\begin{figure}
\centering
\includegraphics[totalheight=1.7in]{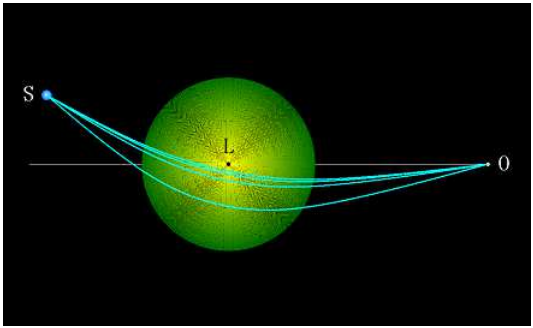}
\caption{\label{fig06} Tracing the central imaging rays for lenses
of different mass.}
\end{figure}

For the actual condition of gravitational imaging, the distances
$OL$, $SL$ and the mass $M$ are all astronomical figures; therefore,
the light of central imaging is extremely close to the center of the
lens. However, for a lensing body with a density increasing towards
the center, it is possible that there are barrier matters near the
center which will destroy the formation of the central image.
Besides, the relatively longer inner path of the central imaging
light adds the possibility of light being held back by the lens
matters on the way. These and some other factors such as the
relative faintness of the central imaging light and the possibly
higher brightness of the lens core itself, all decrease the
possibility of central imaging being actually observed. This
analysis is firmly supported by the fact that the number of observed
images is not ``odd'' as expected by the existed theories but
``even'' in almost all cases of gravitational lensing
\cite{rf-Winn2004}.

\section{Conclusions}

We have proposed the concept of inhomogeneous vacuum with graded
refractive index based on the analysis of the influence of static
gravitational field on the vacuum. we derived the expression of this
refractive index analytically. By using this expression, we
investigated the effect of gravitational lensing in a conventional
optical way and provided a reasonable interpretation for the problem
of central image missing.

The result indicates that, the concept of inhomogeneous vacuum is
mathematically equivalent to the curved spacetime in the general
relativity; therefore, an effective and convenient alternative
method (i.e., optical method) could be established to solve the so
complicated problems in gravitational astronomy. Physically, under
such point of view, the motion of light in gravitational space is a
motion of light wave in a quantum vacuum with graded refractive
index. And as we know that, in conventional optics, the Fermat's
principle says that the optical path between two given points is an
extremum. This is also equivalent to the theorem in the general
relativity that a particle always moves along a geodesic line in a
curved spacetime.


\appendix{\textbf{Acknowledgments}}

We wish to acknowledge the supports from the National Key Project
for Fundamental Research (grant no. 2006CB921403), the National
Hi-tech project (grant no. 2006 AA06A204) and the Zhejiang
Provincial Qian-Jiang-Ren-Cai Project of China (grant no.
2006R10025).


\begin{thebibliography}{00}




\bibitem{rf-Gies2006}
H. Gies and K. Klingm{\"{u}}ller, Phys.\ Rev.\ Lett. \textbf{96},
 220401 (2006).

\bibitem{rf-Emig2006}
T. Emig \emph{et al}., Phys.\ Rev.\ Lett. \textbf{96}, 080403
(2006).

\bibitem{rf-Lamoreaux1997}
S. K. Lamoreaux, Phys.\ Rev.\ Lett. \textbf{78}, 5 (1997).

\bibitem{rf-Chan2001}
H. B. Chan \emph{et al}., Science \textbf{291}, 1941 (2001).

\bibitem{rf-Ahmadi2006}
N. Ahmadi and M. Nouri-Zonoz, Phys.\ Rev.\ D. \textbf{74}, 044034
(2006).

\bibitem{rf-Dupays2005}
A. Dupays \emph{et al}., Phys.\ Rev.\ Lett. \textbf{94}, 161101
(2005).

\bibitem{rf-Rikken2003}
G. L. J. A. Rikken and C. Rizzo, Phys.\ Rev.\ A. \textbf{63}, 012107
(2000) ; \textbf{67}, 015801 (2003).

\bibitem{rf-Armoni2005}
A. Armoni, A. Gorsky, and M. Shifman, Phys.\ Rev.\ D. \textbf{72},
105001 (2005).

\bibitem{rf-Barroso2006}
A. Barroso \emph{et al}., Phys.\ Rev.\ D. \textbf{74}, 085016 (2006)
.

\bibitem{rf-Dienes2005}
K. R. Dienes, E. Dudas, and T. Gherghetta, Phys.\ Rev.\ D.
\textbf{72}, 026005 (2005).

\bibitem{rf-Winn2004}
J. N. Winn, D. Rusin, and C. S. Kochanek, Nature \textbf{427}, 613
(2004).

\bibitem{rf-Landau1975}
L. D. Landau and E. M. Lifshitz, \emph{The Classical Theory of
Fields} (Pergamon Press, New York, 1975).

\bibitem{rf-Wolf1999}
M. Born and E. Wolf, \emph{Principles of Optics} (7th edition)
(Cambridge University Press, Cambridge, 1999).

\bibitem{rf-Einstein1923}
H. A. Lorentz, A. Einstein, and H. Minkowski, \emph{The Principle of
Relativity} (Methuen, London, 1923).

\bibitem{rf-Ohanian1976}
H. C. Ohanian, \emph{Gravitation and Spacetime} (W. W. Norton and
Company Inc., New York, 1976).

\bibitem{rf-Weinberg1972}
S. Weinberg, \emph{Gravitation and Cosmology} (John Wiley and Sons,
New York, 1972).

\bibitem{rf-Fomaleont1976}
E. B. Fomaleont and R. A. Sramek, Phys.\ Rev.\ Lett. \textbf{36},
1475 (1976).

\bibitem{rf-Mollerach2002}
S. Mollerach and E. Roulet, \emph{Gravitational Lensing and
Microlensing} (World Scientific, New Jersey, 2002).




\end{thebibliography}
\end{document}